\documentclass[%
reprint,
superscriptaddress,
amsmath,
amssymb,
aps,
prl,
]{revtex4-2}

\usepackage{graphicx}
\usepackage{subfigure}
\usepackage{verbatim}
\usepackage{color} 
\usepackage{dcolumn}
\usepackage{bm}
\usepackage[colorlinks=true, allcolors=blue]{hyperref}
\begin{document}

\title{Low-Dimensional Phase Diagram of Higher-Order Networked Systems}

\author{Jia-Jie Qin}
\affiliation{School of Physical Science and Engineering, and Center for Frontier Interdisciplinary Research in Fundamental Sciences, Tongji University, Shanghai 200092, P. R. China}

\author{Jack Murdoch Moore}
\affiliation{School of Physical Science and Engineering, and Center for Frontier Interdisciplinary Research in Fundamental Sciences, Tongji University, Shanghai 200092, P. R. China}

\author{Xiaozhu Zhang}\thanks{xiaozhu\underline{ }zhang@tongji.edu.cn}
\affiliation{School of Physical Science and Engineering, and Center for Frontier Interdisciplinary Research in Fundamental Sciences, Tongji University, Shanghai 200092, P. R. China}

\author{Gang Yan}\thanks{gyan@tongji.edu.cn}
\affiliation{School of Physical Science and Engineering, and Center for Frontier Interdisciplinary Research in Fundamental Sciences, Tongji University, Shanghai 200092, P. R. China}
\affiliation{State Key Laboratory of Autonomous Intelligent Unmanned Systems, Tongji University, Shanghai 201210, P. R. China}

\date{\today}

\begin{abstract}
Higher-order networks exhibit rich critical phenomena that cannot be captured by traditional pairwise models. Here, we develop an analytical dimension-reduction framework that maps higher-order networked dynamics onto an effective low-dimensional system, allowing accurate prediction of tipping boundaries, bistability regions, and the nature of phase transitions. We demonstrate the power of this framework across a range of dynamical processes, revealing distinct effects of higher-order interactions on transition continuity and hysteresis. Furthermore, we find that system resilience exhibits a profound dependence on the alignment between pairwise and higher-order connectivity, with assortative mixing enhancing tipping toward active states. Our findings establish a general theory for understanding the critical transitions in higher-order networks, offering new insights for anticipating and managing systemic risk in complex systems.
\end{abstract}

\maketitle

\textit{Introduction} --
Collective behavior in complex networked systems emerges from the dynamics of nodes and interactions, each shaped by a set of governing parameters. When certain parameters cross a critical threshold, known as a tipping point, the system can undergo an abrupt transition, switching from a functioning, active state to complete inactivity, or vice versa~\cite{Dorogovtsev2008RevModPhys1275}. Such shifts underlie phenomena as diverse as epidemic outbreaks~\cite{PastorSatorras2001PhysRevLett3200}, power blackouts~\cite{Buldyrev2010Nature1025}, species extinction~\cite{Scheffer2012Science344}, financial crises~\cite{frank2009science422}, and supply chain disruptions~\cite{Moran2025PNAS2415139122}. Anticipating these transitions is crucial to preserving system resilience, as the capacity to predict an impending tipping point can mark the difference between a recoverable disturbance and irreversible collapse~\cite{scheffer2009nature7260, Masuda2024NatCommun1086, liu2024prx031009}. Identifying these thresholds is a nontrivial task even for systems with traditional dyadic interactions. The large number of nodes, complex connectivity patterns~\cite{reka2002rmp47}, specific forms of governing dynamics~\cite{sanhedrai2022NP338}, and sensitivity to initial conditions~\cite{Taylor2005EcolLett895} combine to obscure the geometry of the resilience landscape.

One powerful strategy for tipping point prediction is \emph{dimensional reduction}, i.e., constructing an effective lower-dimensional system that preserves the essential nonlinear behaviors of the original system while being far easier to analyse. For networked systems with pairwise interactions, several dimensional reduction approaches have been developed in recent years~\cite{gao2016nature307, laurence2019prx011042, Marina2023pnasnex150, wu2023prl097401,morone2019NP95, Vincent2020prr043215, naoki2022prr023257, qin2023prr043209}. A degree-weighted average operator can capture global behaviour~\cite{gao2016nature307}, spectral methods using dominant eigenvectors have proved effective~\cite{laurence2019prx011042, Marina2023pnasnex150}, and recent results have established rigorous bounds on tipping points for nonlinear cooperative systems~\cite{wu2023prl097401}, informed by insights into network core structure~\cite{morone2019NP95}. However, real-world systems often feature interactions beyond simple dyads. Higher-order interactions~\cite{Benson2016Science163, battiston2021np1093, Lambiotte2019NatPhys313}, involving three or more nodes simultaneously, are integral to the dynamics of neural~\cite{chad2015pnas13455, gardner2022nature123}, ecological~\cite{mayfield2017nee0062, cervantes2021el1443, bairey2016nc12285}, and social systems~\cite{cencetti2021sr7028, alvarez-rodriguez2021nhb586}, among others. These multibody interactions can qualitatively alter collective behaviour~\cite{Battiston2026NatRevPhys146, mill2020prl218301, lucas2020prr033140, zhang2023nc1605, neuh2020pre032310, Ji2024pre044204} and even the nature of phase transitions~\cite{Skardal2019PhysRevLett248301, iacopini2019nc2485}. This poses new analytical challenges as most dimensional reduction techniques developed for pairwise networks do not extend naturally to such higher-order systems.

Here we develop a dimension reduction framework capable of predicting tipping points and identifying phase diagrams for networked systems with higher-order interactions. The method introduces a small set of weighted observables, one for each interaction order, that encapsulate the contributions of the corresponding interaction structure. This leads to a reduced system whose dimension equals the highest interaction order present, usually two or three in empirical networks, while preserving the essential nonlinear features of the original dynamics. Applying the framework to epidemic, gene-regulatory, and neuronal dynamics, we demonstrate accurate prediction of phase boundaries and bifurcations. Moreover, we uncover an inter-order structural effect: the alignment between pairwise and higher-order connectivity markedly impact system resilience.

\begin{figure}[t]
    \centering
    \includegraphics[width=\columnwidth]{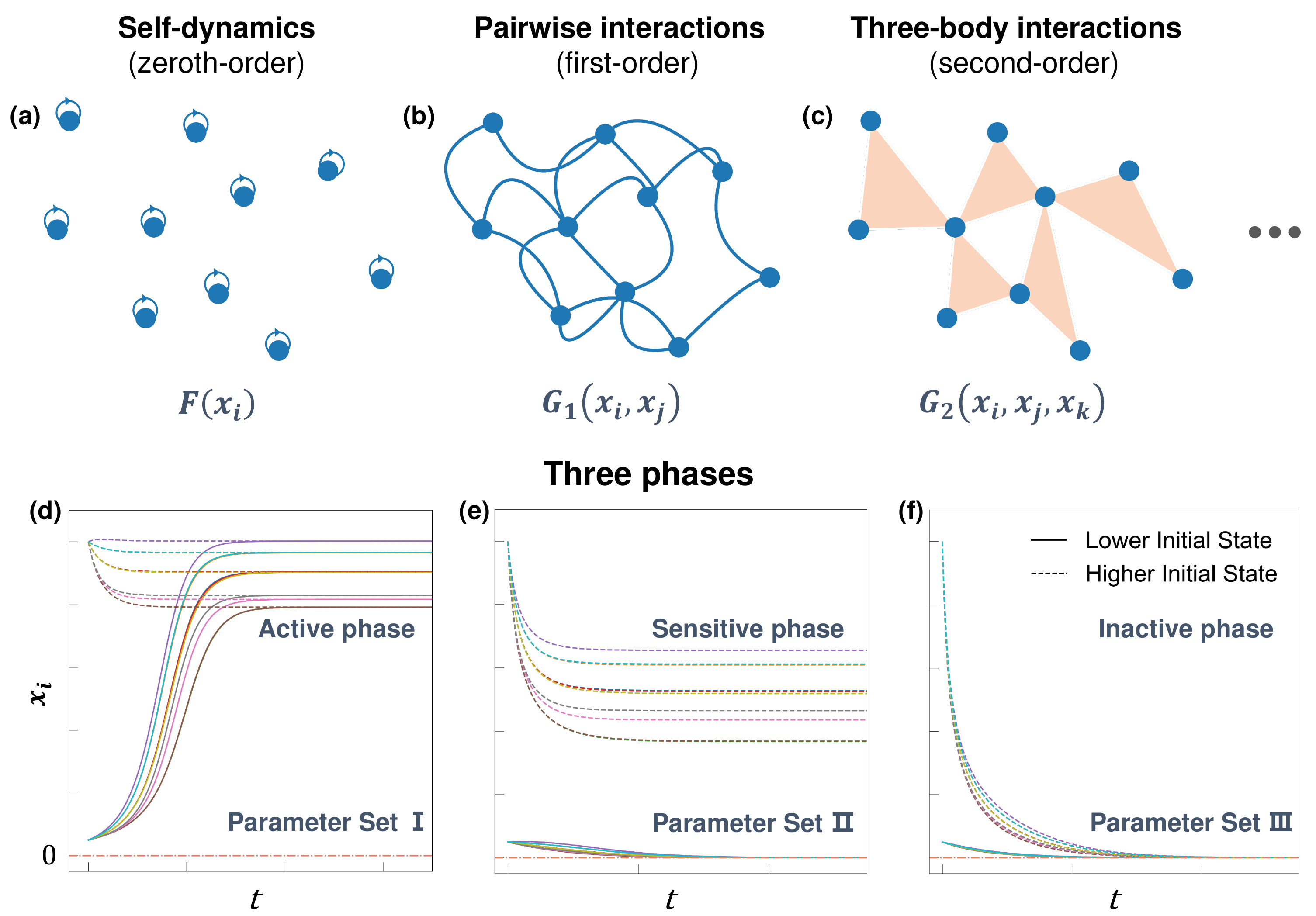}
    \caption{
    \textbf{Higher-order networks and three dynamical phases.}
    (a–c) Interaction orders representing self-dynamics (zeroth order), pairwise (first order) and three-body (second order) couplings. (d–f) Under three representative parameter sets, the system evolves to distinct steady states depending on initial conditions, illustrating active, sensitive, and inactive phases.}
    \label{fig1}
\end{figure}

\textit{Dimension Reduction} --
We illustrate our approach for a system with both pairwise and three-body interactions, the simplest case beyond purely dyadic dynamics. The state of node $i$ is denoted $x_i(t)$ and evolves as
\begin{equation}
    \begin{aligned}
    \dot{x}_i=F(x_i)&+g_1\sum_{j=1}^N A_{ij}G_1(x_i,x_j) \\
    &+\frac{g_2}{2}\sum_{j,k=1}^N B_{ijk}G_2(x_i,x_j,x_k),
    \end{aligned}
    \label{eq:general_dynamics_D2}
\end{equation}
where the terms $F$, $G_1$, and $G_2$ capture the contributions from self-dynamics at the single-node level (zeroth order), pairwise interactions between connected nodes (first order), and three-body interactions among triplets of nodes (second order), respectively (Fig.~\ref{fig1}a–c). The parameters $g_1$ and $g_2/2$ quantify the coupling strengths associated with the latter two interaction orders, while $A_{ij}$ and $B_{ijk}$ are adjacency tensors encoding the presence of edges and triplets. 

Different values of the parameters in Eq.~\eqref{eq:general_dynamics_D2}, particularly the second-order coupling strength $g_2$, can lead the system to exhibit distinct dynamical regions. For instance, under some parameter sets I and III, the system consistently evolves toward an active state (Fig.~\ref{fig1}d) or an inactive state (Fig.~\ref{fig1}f), where the average activity of all nodes remains strictly positive or vanishes, respectively. In contrast, there could exist a parameter set II, within which the final outcome depends sensitively on the initial states: high initial states converge to a non-zero activity, while low initial states decay to zero (Fig.~\ref{fig1}e). This dependence on initial conditions signals the presence of hysteresis in the phase transition, a hallmark phenomenon that is particularly pronounced in higher-order networked systems~\cite{iacopini2019nc2485, ghosh2023chaos053117, li2024amc128832}. This increased complexity implies that, beyond traditional bistability, one must consider the tipping points of multistable regions when characterizing transitions in higher-order networked systems.

Previous dimensional reduction theories, such as the degree-based effective state framework~\cite{gao2016nature307}, typically assume that the macroscopic evolution of a system can be dominated by a single order parameter. However, in systems encompassing higher-order interactions, the tensor $B_{ijk}$ introduces multi-body synergistic effects that are independent of pairwise interactions $A_{ij}$. As a result, the effective input to a node arises not only from pairwise but also from higher-order interactions. Attempting to describe the higher-order system with a single macroscopic variable effectively assumes that these two types of mean fields can be simultaneously represented by the same effective state. This assumption is only approximately valid when the higher-order structure is closely correlated with the pairwise structure~\cite{ghosh2023chaos053117}. In general higher-order networks, these two types of interactions reflect distinct structural features, and a single-variable reduction tends to blend or obscure the individual contributions of each interaction channel.

Therefore, the reduction of higher-order dynamical systems requires identifying distinct effective mean fields associated with different interaction orders. To capture these macroscopic evolution patterns, we introduce two weighted observables
\begin{equation}
x = \sum_{i=1}^N a_i x_i, \quad y = \sum_{i=1}^N b_i x_i,
\label{observable}
\end{equation}
where the weight vectors $\boldsymbol{a}$ and $\boldsymbol{b}$ quantify each node's contribution to the collective dynamics through pairwise and three-body interactions, respectively. We now show how these weight vectors are constrained by the network topology and how they lead to the governing equations of the reduced system.

To this end, we project Eq.~\eqref{eq:general_dynamics_D2} onto the observables defined in Eq.~\eqref{observable}. We then expand $F$ around the macroscopic state and expand $G_1$ and $G_2$ around the linear manifolds induced by the observables (see Supplementary Information Section I). This procedure maps the original $N$-dimensional system onto the following 2-dimensional effective dynamics:
\begin{equation}
    \begin{aligned}
        \dot{x} &\approx F(x) + g_1\alpha _1 G_1(\mu_1 x,\gamma_1 x)  + g_2\beta _1 G_2(\phi_1 y, \omega_1 y, \omega_1 y), \\
    	\dot{y} &\approx F(y) + g_1\alpha _2 G_1(\mu_2 x,\gamma_2 x)  + g_2\beta _2 G_2(\phi_2 y, \omega_2 y, \omega_2 y),
    \end{aligned}
    \label{eq:reduced_dynamics_order}
\end{equation}
where $\mu_\ell, \gamma_\ell, \phi_\ell, \omega_\ell$ $(\ell=1,2)$ are structural descriptors determined by the compatibility equations
\begin{equation}
    \begin{aligned}
    &\boldsymbol{D}_1 \boldsymbol{a}_\ell = \alpha_\ell \mu_\ell \boldsymbol{a}, 
    \quad \boldsymbol{A} \boldsymbol{a}_\ell = \alpha_\ell \gamma_\ell \bm{a},\\
    &\boldsymbol{D}_2 \boldsymbol{b} = \beta_\ell \phi_\ell \boldsymbol{a}_\ell,  
    \quad \tilde{\boldsymbol{B}} \boldsymbol{a}_\ell = \beta_\ell \omega_\ell \bm{b}.
    \end{aligned}
     \label{eq:structural compatibility equations}
\end{equation}
where $\tilde{\bm{B}}$ is defined as $\tilde{B}_{ij} = \sum_{k=1}^N B_{ijk}/2$, encoding the three-body interaction structure; $\bm{D}_1$ and $\bm{D}_2$ are diagonal matrices with entries $D_1(i,i) = d_i^{(1)}$ and $D_2(i,i) = d_i^{(2)}$, where $d_{i}^{(1)} = \sum_{j=1}^N A_{ij}$ and $d_{i}^{(2)} = \sum_{j=1}^N \tilde{B}_{ij}$ denote the degree and hyperdegree of node $i$, respectively; The shorthand $\{\bm{a}_\ell|\ell=1,2\}$ denotes the weight vectors $\{\bm{a},\bm{b}\}$, while $\alpha_\ell=\bm{a}_\ell^T\bm{d}^{(1)}$ and $\beta_\ell=\bm{a}_\ell^T\bm{d}^{(2)}$ represent the degree and hyperdegree weighted by $\bm{a}$ and $\bm{b}$, respectively.

\begin{figure*}[ht]
    \centering
    \includegraphics[width=0.7\textwidth]{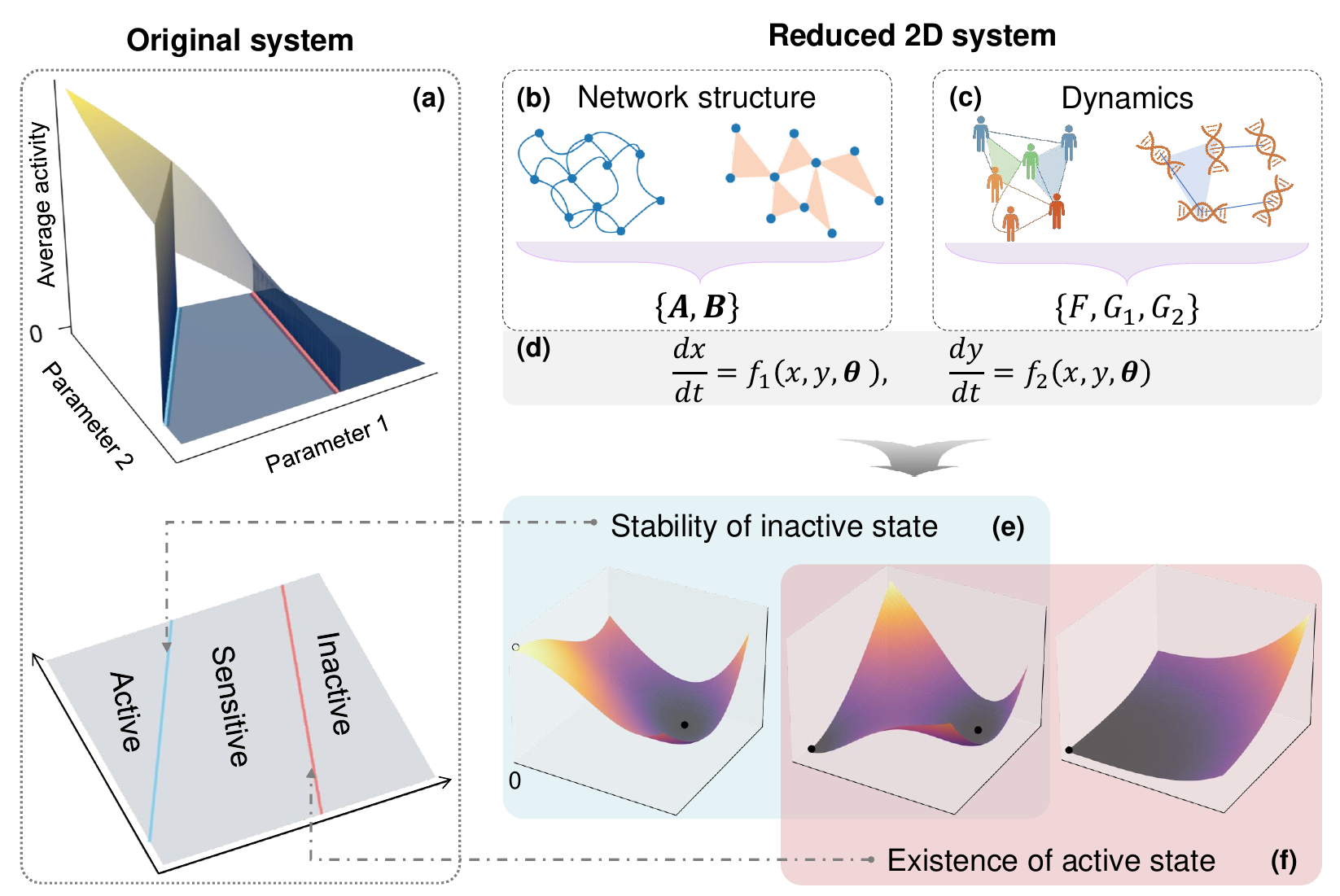}
    \caption{
    \textbf{Dimension reduction framework for higher-order networked systems.} Starting from a higher-order network (b) and the associated dynamical process (c), our framework reduces the networked dynamics to an effective two-dimensional system (d). This reduction allows to analyze the stability of inactive state (e) and the existence of active state (f), determing the boundaries among active, sensitive, and inactive phases thereby yielding the full phase diagram (a).}
    \label{fig:bifurcation_diagram}
\end{figure*}

The remaining task is to determine the structural descriptors \(\mu_\ell, \gamma_\ell, \phi_\ell, \omega_\ell\) by solving Eq.~\eqref{eq:structural compatibility equations}. Leveraging the convexity of the problem, we adopt the residual minimization approach of Ref.~\cite{laurence2019prx011042}, which yields explicit expressions for the descriptors in terms of the weight vectors \(\bm{a}\) and \(\bm{b}\) (see Supplementary Information Section I for detailed derivation). Recognizing that the dominant eigenvectors generally capture the leading influence of network structure on system dynamics, we choose the normalized dominant eigenvectors of \(\boldsymbol{A}\) and \(\tilde{\boldsymbol{B}}\) to specify the weight vectors, yielding
\begin{equation}
    \begin{aligned}
    &\mu_1 = \dfrac{{\boldsymbol{a}}^T \boldsymbol{D}_1 {\boldsymbol{a}}}{\alpha_1 {\boldsymbol{a}}^T {\boldsymbol{a}}}, \gamma_1=1, 
    \phi_1 = \dfrac{{\boldsymbol{b}}^T \boldsymbol{D}_2 {\boldsymbol{a}}}{\beta_1 {\boldsymbol{b}}^T {\boldsymbol{b}}}, \omega_1= \dfrac{{\boldsymbol{b}}^T \tilde{\boldsymbol{B}} {\boldsymbol{a}}}{\beta_1 {\boldsymbol{b}}^T {\boldsymbol{b}}}\\
    &\mu_2 = \dfrac{{\boldsymbol{a}}^T \boldsymbol{D}_1 {\boldsymbol{b}}}{\alpha_2 {\boldsymbol{a}}^T {\boldsymbol{a}}}, \gamma_2 = \dfrac{{\boldsymbol{a}}^T \boldsymbol{A} {\boldsymbol{b}}}{\alpha_2 {\boldsymbol{a}}^T {\boldsymbol{a}}},
    \phi_2 = \dfrac{{\boldsymbol{b}}^T \boldsymbol{D}_2 {\boldsymbol{b}}}{\beta_2 {\boldsymbol{b}}^T {\boldsymbol{b}}}, \omega_2= 1. 
\end{aligned}
\label{eq:coeff_order}
\end{equation}

\textit{Identifying Phase Diagram} --
The original system in Eq.~\eqref{eq:general_dynamics_D2} can exhibit three distinct dynamical phases—active, sensitive, or inactive—depending on the parameter regime, as illustrated in Fig.~\ref{fig:bifurcation_diagram}a.  Given any higher-order network characterized by $A_{ij}$ and $B_{ijk}$ (Fig.~\ref{fig:bifurcation_diagram}b) and dynamics described by $F$, $G_1$ and $G_2$ (Fig.~\ref{fig:bifurcation_diagram}c), our dimension-reduction framework first computes all relevant structural descriptors via Eq.~\eqref{eq:coeff_order}, and then fully specifies the reduced system Eq.~\eqref{eq:reduced_dynamics_order} (Fig.~\ref{fig:bifurcation_diagram}d). In the following, we validate our framework across various dynamical models and network topologies, demonstrating that the resulting two-dimensional flow allows for efficient determination of the phase boundaries and the nature of the associated phase transitions.

We first consider the Susceptible-Infected-Susceptible (SIS) epidemic spreading dynamics as a representative example, in which the state $x_i$ of each node $i$ evolves according to
\begin{equation}
\begin{aligned}
\dot{x}_i = &\ -s x_i + g_1 \sum_{j} A_{ij} (1 - x_i)x_j \\
&+ \frac{g_2}{2} \sum_{j,k} B_{ijk} (1 - x_i)x_j x_k,
\end{aligned}
\label{eq:sis_dynamics}
\end{equation}
where $x_i$ denotes the infection probability of node $i$. The first term, $F(x_i) = -s x_i$, represents the self-dynamics governed by the recovery rate $s$. The second term, $G_1(x_i,x_j) = (1 - x_i)x_j$, captures pairwise infection through the adjacency matrix $A_{ij}$, with strength $g_1$. The third term, $G_2(x_i,x_j,x_k) = (1 - x_i)x_j x_k$, describes three-body interactions encoded in the tensor $B_{ijk}$, with strength $g_2/2$. Together, these components characterize the combined influence of recovery, pairwise transmission, and higher-order contagion on the epidemic dynamics.
\begin{figure*}[ht]
    \centering
    \includegraphics[scale=0.48]{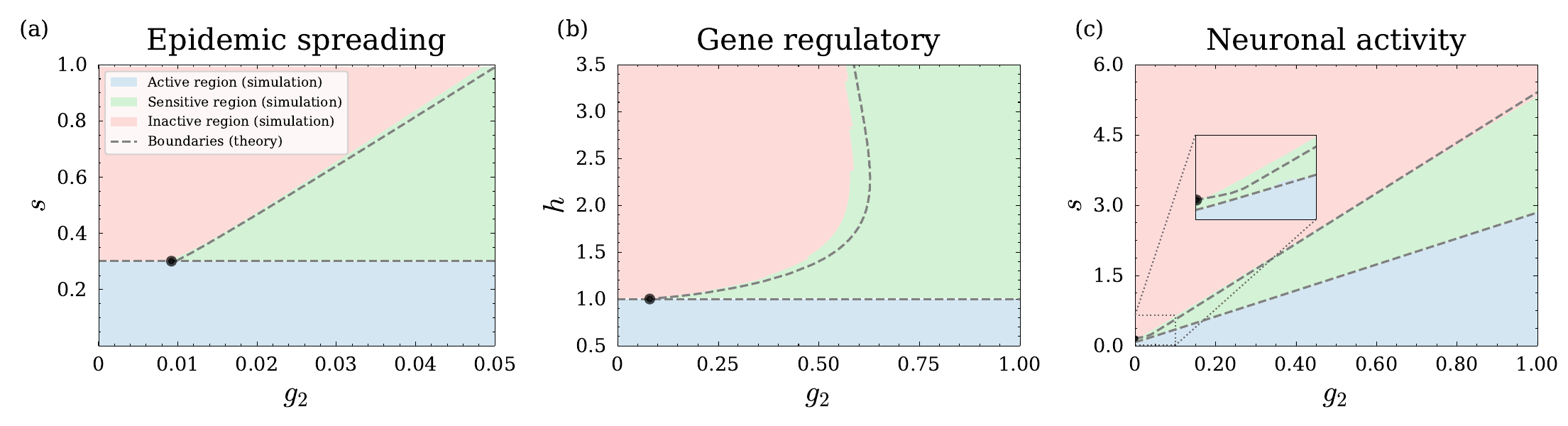}
    \caption{
    \textbf{Phase diagrams for three representative dynamics on higher-order networks.} 
    The inactive (red), sensitive (green), and active (blue) regions obtained from numerical simulations for 
    (a) epidemic spreading dynamics on a high-school contact network ($N=180$ nodes, $L_1=2,220$ edges, $L_2=3,000$ triangles)~\cite{benson2018simplicial} with $g_1 = 0.01$; 
    (b) gene-regulatory dynamics on the yeast protein–protein interaction network ($N=1,647$ nodes, $L_1=5,036$ edges, $L_2=212$ triangles)~\cite{yu2008science104} with $a=1$, $g_1 = 0.5$, $s=7$; 
    (c) neuronal dynamics on a brain white-matter connectivity network ($N=906$ nodes, $L_1=2,225$ edges, $L_2=1,273$ triangles)~\cite{bullmore2009nrn186} with $\delta=1$, $\zeta = 2$, $g_1 = 0.1$. 
    Dashed lines represent the theoretical prediction of phase boundaries from the reduced system Eq.~\eqref{eq:reduced_dynamics_order}, and black dots represent the critical point for the onset of bistability.
    }
    \label{phase diagram}
\end{figure*}

Given a network topology that encodes both pairwise and three-body interaction structures, $A_{ij}$ and $B_{ijk}$, one can construct the matrices $\boldsymbol{A}$ and $\tilde{\boldsymbol{B}}$ and calculate their respective leading eigenvectors, $\boldsymbol{a}$ and $\boldsymbol{b}$. By further extracting the degree and hyperdegree diagonal matrices, $\bm{D}_1$ and $\bm{D}_2$, we determine the values of $\alpha_\ell$ and $\beta_\ell$ $(\ell = 1,2)$. According to Eq.~\eqref{eq:coeff_order}, this procedure yields all eight structural descriptors, which are then substituted into Eq.~\eqref{eq:reduced_dynamics_order}. Consequently, we obtain the reduced two-dimensional flow capturing the epidemic spreading on higher-order networks
\begin{equation}
    \begin{aligned}
        \dot{x} \approx & -s x + g_1 \alpha_1 \big(1 - \mu_1 x\big)x + g_2 \beta_1 \big(1 - \phi_1 y \big)\omega_1^2 y^2,\\
        \dot{y}\approx& -s y + g_1 \alpha_2 \big(1 - \mu_2 x \big)\gamma_2 x+ g_2 \beta_2 \big(1 - \phi_2 y \big)y^2.
    \end{aligned}
    \label{eq:reduced_sis}
\end{equation}

In many empirical systems, spontaneous activity decays in the absence of sufficiently strong interactions. For the epidemic spreading process considered here, the inactive state, where all node activities vanish, therefore constitutes a natural attractor. As the recovery rate $s$  decreases or the interaction strengths $g_1$ and $g_2$ increase, a nontrivial positive fixed point may emerge, causing the inactive state $(x, y) = (0, 0)$ to lose stability (Fig.~\ref{fig:bifurcation_diagram}e). In this region, even small perturbations, such as a few infected nodes, can trigger a macroscopic outbreak. The boundary separating the sensitive and active phases is thus determined by the linear stability of the inactive state. For the SIS dynamics, the leading eigenvalue of the Jacobian matrix of the reduced system Eq.~\eqref{eq:reduced_sis} at the origin is $-s + g_1 \alpha_1$, where $\alpha_1$ is equal to the leading eigenvalue of the adjacency matrix $\bm{A}$. Hence, the active phase emerges when $g_1 > s/\alpha_1$.

In contrast, the boundary between the sensitive and inactive phases is governed by the existence of stable positive fixed point (Fig.~\ref{fig:bifurcation_diagram}f). By solving Eq.~\eqref{eq:reduced_sis} from high initial conditions, we determine this boundary and obtain excellent agreement with direct simulations of SIS dynamics (gray dashed line in Fig.~\ref{phase diagram}a). These results confirm that higher-order interactions can generate bistability and hysteresis, consistent with previous findings~\cite{iacopini2019nc2485}.

A key question, however, is whether bistability arises solely from the presence of higher-order structures or requires sufficiently strong higher-order interactions. To answer this, we analyze the onset of bistability near the active-phase threshold $s\approx g_1\alpha_1$. Assuming weak parameter heterogeneity, we introduce the averaged parameters $\bar{\mu}=(\mu_1+\mu_2)/2$ and $\bar{\phi}=(\phi_1+\phi_2)/2$, and derive an effective two-dimensional description Eq.~(S20) in Supplementary Information. Analysis of its nontrivial fixed points reveals that a discontinuous transition emerges only when the higher-order interaction strength exceeds the critical threshold
\begin{equation}
\label{eq:critical_g2}
g_{2}^{\text{crit}} = \frac{g_1\alpha_1\bar{\mu}}{\beta_1\omega_{1}^{2}}\cdot \frac{4z_*-3c_1}{z_{*}^{2}\left( 2z_*-c_1 \right)},
\end{equation}
where $z_*$ and $c_1$ depend solely on network topology (see Supplementary Information Section IV-A). The theoretical prediction Eq.~\eqref{eq:critical_g2} agrees remarkably well with direct numerical simulations, as indicated by the black dot in Fig.~\ref{phase diagram}a. This agreement further validates our reduced-dimensional analysis and establishes an explicit criterion for when higher-order interactions qualitatively alter the nature of phase transition and induce bistability.

We next consider gene regulatory dynamics on a yeast protein-protein interaction network~\cite{yu2008science104}, with the mathematical expression of self- and interacting dynamics given by Eq.~(S15) in Supplementary Information. Applying our dimension-reduction framework yields the effective two-dimensional system
\begin{equation}
    \begin{aligned}
    \dot{x}\approx & -sx^a+g_1\alpha_1\frac{x^h}{1+x^h}+g_2\beta_1\frac{(2\omega_1y)^h}{1+(2\omega_1y)^h},\\
    \dot{y}\approx & -sy^a+g_1\alpha_2\frac{(\gamma_2x)^h}{1+(\gamma_2x)^h}+g_2\beta_2\frac{(2y)^h}{1+(2y)^h}.
    \end{aligned}
\label{eq:reduced_grn}
\end{equation}

Linear stability analysis shows that the origin is stable for $h<a\le 1$ and unstable otherwise. For the biologically relevant case $a=1$, the active-phase boundary is therefore determined by $h=1$. To locate the boundary between the sensitive and inactive phases, we numerically solve Eq.~\eqref{eq:reduced_grn} from high initial conditions. The resulting prediction agrees closely with direct simulations, as shown in Fig.~\ref{phase diagram}b.

To determine the onset of bistability, we consider the limit $x, y \to 0^+$ and find that a nonzero state exists when the determinant of a particular two-dimensional matrix vanishes (see Supplementary Information Section IV-B), leading to the approximate critical threshold
\begin{equation}
    g_{2}^{\mathrm{crit}} = \frac{s\left(s-g_{1}\alpha_{1}\right)}{\left(s-g_{1}\alpha_{1}\right)\beta_{2}2^{a}+g_{1}\alpha_{2}\beta_{1}\gamma_{2}^{a}(2\omega_{1})^{a}}.
\end{equation}
This theoretical prediction is in excellent agreement with direct numerical simulations (Fig.~\ref{phase diagram}b) and reveals a mechanism distinct from SIS dynamics: if pairwise interactions alone are sufficient to overcome local decay near the inactive state, i.e., when $g_1 \ge s/\alpha_{1}$, bistability can emerge without higher-order reinforcement, corresponding to $g_{2}^{\mathrm{crit}} \le 0$. Otherwise, sustaining an active state requires additional nonlinear feedback from higher-order interactions, and bistability arises only when $g_{2} \ge g_{2}^{\mathrm{crit}}$.


Finally, we consider neuronal dynamics on a structural brain network comprising $998$ regions and $1,273$ triangle-based higher-order interactions~\cite{bullmore2009nrn186}, with self- and interacting forms given by Eq.~(S16) in Supplementary Information. Applying the same reduction procedure yields the effective two-dimensional flow
\begin{equation}
    \begin{aligned}
        \dot{x}&\approx -sx + g_1 \alpha_1 \tau(x) + g_2\beta_1 \tau(2\omega_1 y) ,\\
        \dot{y}&\approx -sy + g_1 \alpha_2 \tau(\gamma_2x) + g_2\beta_2 \tau(2 y),
    \end{aligned}
    \label{eq:reduced_neuronal}
\end{equation}
where $\tau(x) = \frac{1}{1 + e^{\varsigma - \delta x}} - \frac{1}{1 + e^{\varsigma}}$. The stability analysis around $(x, y) = (0, 0)$ shows that the active phase emerges when $s < c(g_1\alpha_1 + g_2\beta_2)$ with constant $c = \frac{\delta e^\varsigma}{(1+e^\varsigma)^2}$ (see Supplementary Information Section IV-C).

The neuronal dynamics can also exhibit bistability even in the absence of higher-order interactions. To demonstrate this, we consider the purely pairwise case $g_2=0$ and find that a positive fixed point exists whenever $s$ is smaller than the critical value
\begin{equation}
    s_{\mathrm{crit}} = g_{1}\alpha_{1}\kappa,
    \label{eq:neuronal_critical}
\end{equation}
where $\kappa = \max_{x>0}\tau(x)/x$. Because $\tau^{\prime}(0)$ is positive and constantly smaller than $\kappa$, there always exists an intermediate parameter regime $s \in (s_0, s_{\mathrm{crit}}]$, with $s_0=g_{1}\alpha_{1}\tau^{\prime}(0)$, in which both the inactive and active states are stable (Fig.~\ref{phase diagram}c inset). This finding demonstrates that bistability is an intrinsic property of the neuronal response function $\tau(x)$ itself, independent of the strength of the first or second order interactions. Higher-order interactions can enlarge the bistable region, but are not required for its existence. 

Taken together, these three representative cases demonstrate that our framework offers a general theory for critical transitions in higher-order networked systems and reveals diverse effects of higher-order interactions on the nature of phase transitions. For SIS dynamics, increasing the higher-order interaction strength $g_2$ can shift the transition from continuous to discontinuous; For gene-regulatory dynamics, the alteration of its transition type occurs only when pairwise interactions are weak; For neuronal dynamics, the transition remains always discontinuous, with $g_2$ affecting only the extent of the bistable region rather than the transition type.

\begin{figure*}[ht]
    \centering
    \includegraphics[width=\textwidth]{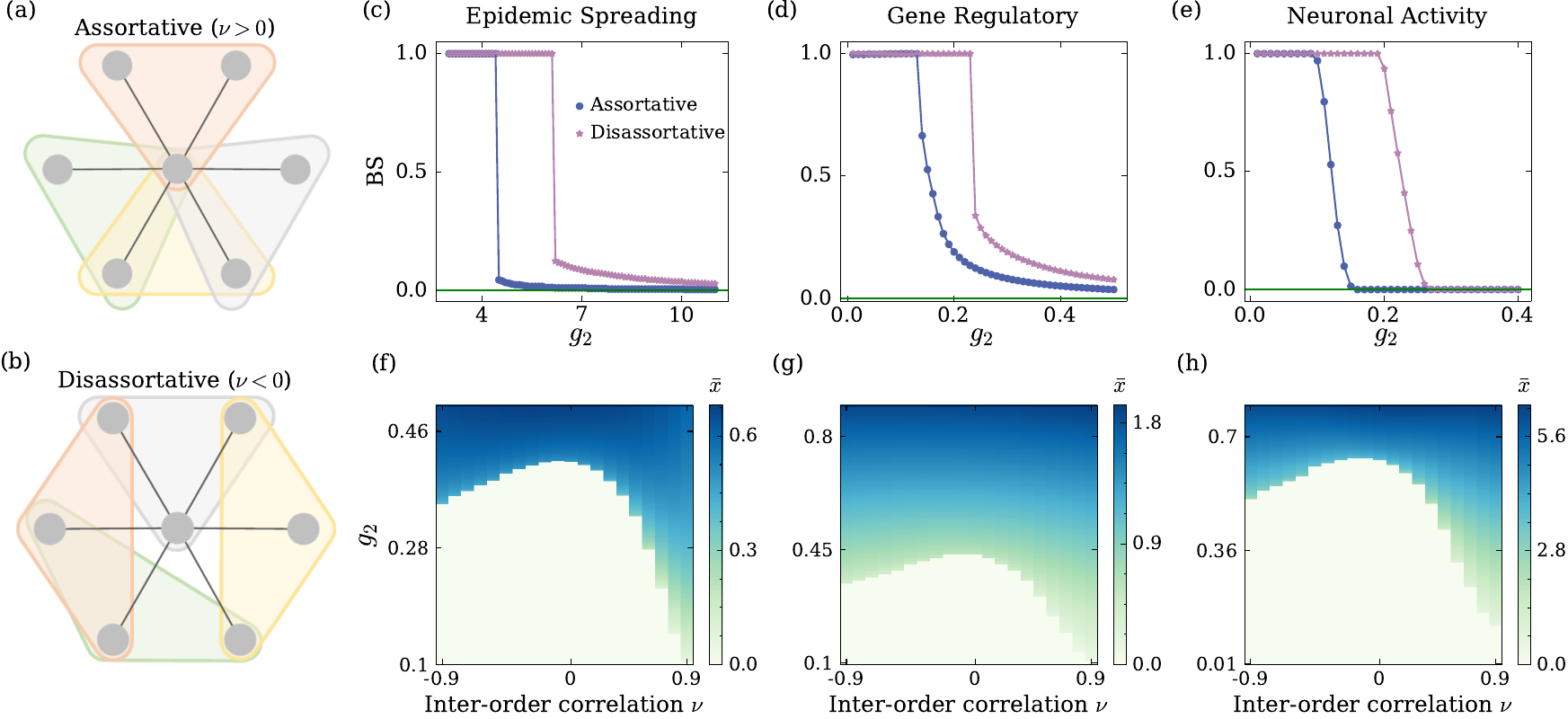}
    \caption{
    \textbf{Effect of inter-order mixing on system resilience.}
    (a,b) Schematic illustration of assortative ($\nu>0$) and disassortative ($\nu<0$) mixing between pairwise and higher-order interactions. (c–e) Basin stability of the inactive state as a function of higher-order interaction strength $g_2$ for both mixing types. (f–h) Average steady-state activity $\bar{x}=\sum_i x_i/N$ as a function of the inter-order correlation $\nu$ and higher-order interaction strength $g_2$, revealing the parameter regimes where the active state persists.}
    \label{bs}
\end{figure*}

\textit{Inter-order correlations and resilience}—
Beyond the abundance of higher-order interactions, the way different interaction orders are correlated can influence system resilience. To quantify this effect, we introduce an inter-order correlation measure based on the rank correspondence between node degree and hyperdegree. Specifically, we define the correlation exponent $\nu$ through
\begin{equation}
d^{(2)} \sim (d^{(1)})^\nu,
\end{equation}
where $d^{(1)}$ and $d^{(2)}$ denote the degree and hyperdegree of a node, respectively. Positive values of $\nu$ indicate assortative coupling, whereby highly connected nodes also participate in many higher-order interactions (Fig.~\ref{bs}a), whereas negative values correspond to disassortative organization (Fig.~\ref{bs}b).

To assess the impact of $\nu$ on tipping dynamics, we compute the basin stability (BS)~\cite{Menck2013NatPhys89} of the inactive state using the reduced two-dimensional system. Specifically, we sample $10^4$ initial conditions uniformly from $[0,1]\times[0,1]$, and BS is defined as the fraction of trajectories that return to the origin. This quantity provides a direct measure of the robustness of the inactive state against finite perturbations.

 As shown in Fig.~\ref{bs}c–e, system resilience exhibits a pronounced dependence on inter-order organization. An increase in the higher-order interaction strength $g_2$ generally diminishes the basin stability of the inactive state, an effect that is further intensified by assortative mixing ($\nu>0$). In contrast, disassortative mixing counteracts this effect and effectively suppresses tipping toward active states. Notably, as shown in Fig.~\ref{bs}f–h, resilience varies non-monotonically with $\nu$: systems with random inter-order mixing (i.e., $\nu\approx 0$) require the strongest higher-order interactions to sustain the active state. Thus, system resilience is shaped not only by the magnitude of higher-order interactions, but also by how these interactions are organized across different orders.

\textit{Discussion and conclusion}--
Our results establish a dimension-reduction framework for anticipating critical transitions in networked dynamical systems with higher-order interactions. By mapping complex networked dynamics onto a low-dimensional representation, the method not only predicts tipping boundaries but also elucidates the mechanisms through which higher-order interactions shape system resilience, multistability, and the nature of phase transitions. Our analysis of inter-order correlations reveals that higher-order structure can either amplify or buffer systemic risk, depending on its alignment with pairwise connectivity.

While our primary analysis focuses on second-order (three-body) interactions—the most ubiquitous and empirically accessible form of non-pairwise coupling—the framework is naturally extensible to even higher orders. As demonstrated in Supplementary Information Fig.~S3, the formalism remains valid for the third-order SIS system, underscoring its generalization capability.

Beyond its theoretical utility, our approach can be integrated with recent advances in dynamical inference~\cite{Gao2022NatComputSci160, Gao2024NatCommun6029, Hu2025NatCommun6226, Yu2026NatComputSci156}, which enable full reconstruction of the functional forms of self-dynamics $F$, pairwise interactions $G_1$, and second-order interactions $G_2$ from observational data. Merging such data-driven reconstruction with our analytical framework offers a promising paradigm for probing unmeasured collective dynamics in real-world higher-order networks, even when prior mechanistic knowledge is limited.
\bibliography{apssamp}
\end{document}